\documentclass[5p]{elsarticle}

\usepackage{amssymb}
\usepackage{amsfonts}
\usepackage{amsmath}

\usepackage{multirow}
\usepackage{subfig}
\usepackage{float}
\usepackage{caption}
\usepackage{graphicx}
\usepackage{hyperref}

\hypersetup{
    colorlinks,
    linkcolor={red!50!black},
    citecolor={blue!50!black},
    urlcolor={blue!80!black}
}

\def\x{{\mathbf x}}

\newcommand{\fvec}[1]{\pmb{#1}}

\newcommand{\Figref}[1]{Fig.~\ref{#1}}
\newcommand{\Secref}[1]{Section~\ref{#1}}

\newcommand{\Subsubsecref}[1]{Section~\ref{#1}}
\newcommand{\Eqref}[1]{Eq.~\ref{#1}}
\newcommand{\Tabref}[1]{Table~\ref{#1}}

\journal{Journal of the Franklin Institute}

\begin{document}

\begin{frontmatter}

\title{AudioMNIST: Exploring Explainable Artificial Intelligence\\ for Audio Analysis on a Simple Benchmark}

\author[hhi-ai]{Sören Becker$^*$}
\author[hhi-ai]{Johanna Vielhaben$^*$}
\author[hhi-ai]{Marcel Ackermann}
\author[tub-cs,kor,mpi]{Klaus-Robert Müller}
\author[hhi-ai]{Sebastian Lapuschkin$^{**}$}
\author[hhi-ai,tub-cs,bifold]{Wojciech Samek$^{**}$}

\address[hhi-ai]{Department of Artificial Intelligence, Fraunhofer Heinrich-Hertz-Institute, Berlin, Germany}
            
\address[tub-cs]{Department of Electrical Engineering and Computer Science, Technische Universität Berlin, Berlin, Germany}

\address[kor]{Department of Brain and Cognitive Engineering, Korea University, Seoul, Korea}

\address[mpi]{Max Planck Institute for Informatics, Saarbrücken, Germany}

\address[bifold]{BIFOLD – Berlin Institute for the Foundations of Learning and Data, Berlin, Germany}

  \cortext[cor1]{equal contribution}
   \cortext[cor1]{corresponding author}

\begin{abstract}
Explainable Artificial Intelligence (XAI) is targeted at understanding how models perform feature selection and derive their classification decisions.
This paper explores post-hoc explanations for deep neural networks in the audio domain.
Notably, we present a novel Open Source audio dataset consisting of 30,000 audio samples of English spoken digits which we use for classification tasks on spoken digits and speakers' biological sex.
We use the popular XAI technique Layer-wise Relevance Propagation (LRP) to identify relevant features for two neural network architectures that process either waveform or spectrogram representations of the data.
Based on the relevance scores obtained from LRP, hypotheses about the neural networks' feature selection are derived and subsequently tested through systematic manipulations of the input data.
Further, we take a step beyond visual explanations and introduce audible heatmaps. We demonstrate the superior interpretability of audible explanations over visual ones  in a human user study. 
\end{abstract}


\end{frontmatter}

\section{Introduction}
\label{sec:intro}
Deep neural networks, owing to their intricate and non-linear hierarchical architecture, are widely regarded as black boxes with regard to the complex connection between input data and the resultant network output. 
This lack of transparency not only poses a significant challenge for researchers and engineers engaged in the utilization of these models but also renders them entirely unsuitable for domains where understanding and verification of predictions are indispensable, such as healthcare applications \cite{caruana2015intelligible}.

In response, the field of Explainable Artificial Intelligence (XAI) investigates methods to make the classification strategies of complex models comprehensible, including methods introspecting learned features \cite{hinton2006unsupervised,erhan2009visualizing} and methods explaining model decisions
\cite{baehrens2010explain, bach2015pixel,
fong2017interpretable, MonPR17,samek2021explaining}.  The latter methods were initially successful in the realm of image classifiers and have more recently been adapted for other domains, such as natural language processing \cite{ArrWASSA17}, physiological signals \cite{StuJNM16, Strodthoff2018DetectingAI}, medical imaging \cite{THOMAS2022972, klauschen2024pathology}, and physics \cite{schutt2017quantum, bluecher2020ft}.

This paper explores deep neural network interpretation in the audio domain. As in the visual domain, neural networks have fostered progress in audio processing \cite{lee2009unsupervised,hinton2012deep,
2016arXiv161000087D}, particularly in automatic speech recognition (ASR) \cite{rabiner1993fundamentals,anusuya2010speech}. While large corpora of annotated speech data are available (e.g.\ \cite{godfrey1992switchboard,garofolo1993darpa,panayotov2015librispeech}), this paper introduces a simple publicly available dataset of spoken digits in English. The purpose of this dataset is to serve as a basic classification benchmark for evaluating novel model architectures and XAI algorithms in the audio domain. 
Drawing inspiration from the influential MNIST dataset of handwritten digits \cite{lecun1998mnist} that has played a pivotal role in computer vision, we have named our novel dataset \mbox{\emph{AudioMNIST}} to highlight its conceptual similarity.
AudioMNIST allows for several different classification tasks of which we explore spoken digit recognition and recognition of speakers' sex. In this work, we train two neural networks for each task on two different audio representations. Specifically, we train one model on the time-frequency spectrogram representations of the audio recordings and another one directly on the raw waveform representation. We then use a popular post-hoc XAI method called layer-wise relevance propagation (LRP) \cite{bach2015pixel} to inspect the which features in the input are influential for the final model prediction. From these we can derive insights about the model's high-level classification strategies and demonstrate that the spectrogram-based sex classification is mainly based on differences in lower frequency ranges and that models trained on raw waveforms focus on a rather small fraction of the input data.
Further, we explore explanation formats beyond visualization of relevance heatmaps that indicate the impact of each timepoint in the raw waveform or time-frequency component in the spectrogram representation towards the model prediction. Notably, we introduce audible heatmaps and demonstrate their superior interpretability over visual explanations in the audio domain in a human user-study.

The structure of the paper is as follows: In \Secref{sec:methods}, we outline the audio representations used for neural network models, examine LRP as an approach for explaining classifier decisions, and introduce audible explanations generated from LRP relevances while contrasting them with visual explanations. Moving to \Secref{sec:results}, we present the AudioMNIST dataset and delve into the visual explanations derived from LRP. Subsequently, we evaluate the interpretability of these visual explanations compared to audible explanations for practitioners in a human user study.

\section{Explainable AI in the audio domain}
\label{sec:methods}
In this section, we provide an overview of audio representations utilized in the neural network models for audio. We also explore LRP as a method for explaining the output of these models. Finally, we present audible explanations generated from LRP relevances and compare them to visual explanations.

\subsection{Audio representations for NN models}

In the realm of audio signal processing, the raw waveform and the spectrogram serve as fundamental representation formats for neural network-based models. These formats bear striking similarities to images in computer vision, exhibiting translation invariance and sparse unstructured data in either 1-dimensional or 2-dimensional form. Consequently, conventional convolutional neural networks (CNNs) used in computer vision can be trained on these audio signals \cite{2016arXiv160909430H,2016arXiv161000087D}.

\paragraph{Waveform} Straightforwardly, an audio signal in the time domain is represented by a waveform $\fvec{x} \in \mathbb{R}^{L}$ which contains the amplitude values $x_t$ of the signal over time. The time steps between the signal values are determined by the sampling frequency $f_S$, and the duration of the signal is $\frac{L}{f_S}$.

\paragraph{Spectrogram} Alternatively, we can represent the audio signal in the time-frequency domain. Short Time Discrete Fourier Transform (STDFT) transforms the raw  waveform $\fvec{x}$ to its representation  $\fvec{Y}$ in time-frequency domain, and is defined as, 
\begin{equation}
     Y_{k,m} = \sum_{n=0}^{N-1}x_{n+mH} \cdot w_{n} \cdot e^{-\frac{i \pi kn}{N}} \, .
    \label{eq:stdft}
\end{equation}
The STDFT calculates a Discrete Fourier Transform for overlapping windowed parts of the signal. The window function $w$ has a length of $M$ and a hop size of $H$. The resulting spectrogram $\fvec{Y} \in \mathbb{C}^{(K+1) \times M}$ contains complex-valued time-frequency components in $K+1$ frequency bins $k$ and $M$ time bins $m$, where $K=\frac{N}{2}$ and $M = \frac{L-N}{H}$. Usually, the phase information is disregarded, and only the amplitude of the complex spectrogram $\fvec{Y}_{\text{magn}} \in \mathbb{R}^{K+1 \times M}$ is considered for training classifiers.

\subsection{Post-hoc explainability via \\ Layer-wise Relevance Propagation}
CNNs for audio processing are inherently black-box models. In order to understand their inner workings and classfication strategies, we can emply post-hoc explanation methods, see \cite{samek2021explaining, SamXAI19} for a recent overview of current approaches. Here, we focus on a popular method called \emph{Layer-wise Relevance Propagation} (LRP)  \cite{bach2015pixel}, which has been successfully applied to time series data in previous studies \cite{StuJNM16, strodthoff_ecg_xai, Strodthoff2018DetectingAI, gait_2022}. LRP allows for a decomposition of a learned non-linear predictor output $f(\fvec{x})$ into relevance values $R_i$ that are associated with the components $i$ of input $\fvec{x}$. Starting with the output, LRP performs per-neuron decompositions in a top-down manner by iterating over all layers of the network and propagating relevance scores $R_i$ from neurons of hidden layers step-by-step towards the input.
Each $R_i$ describes the contribution an input or hidden variable $x_i$ has made to the final prediction.
The core of the method is the redistribution of the relevance value $R_j$ of an upper layer neuron towards the layer inputs $x_i$, in proportion to the contribution of each input to the activation of the output neuron $j$.
\begin{equation}
R_{i\leftarrow j} = \frac{z_{ij}}{\sum_i z_{ij}} R_j
\label{eq:lrpdecompose}
\end{equation}
The variable $z_{ij}$ describes the forward contribution (or pre-activation) sent from input $i$ to output $j$.
The relevance score $R_i$ at neuron $i$ is then obtained by pooling all incoming relevance quantities $R_{i\leftarrow j}$ from neurons $j$ to which $i$ contributes:
\begin{equation}
R_i = \sum_j R_{i \leftarrow j}
\label{eq:lrppool}
\end{equation}
The initial relevance value equals the activation of the output neuron, for deeper layers it is specified by the choice of redistribution rule depending on the layer's type and position in the model \cite{lapuschkin2019unmasking,kohlbrenner2020towards,samek2021explaining}.
Implementations of the algorithm are publicly available \cite{LapJMLR16,Alber2018iNNvestigateNN, Anders2021SoftwareFD}.

\subsection{Explanation formats}
We can easily employ LRP to deep audio classification models trained on the raw waveform $\fvec{x}$ or the spectrogram representation $\fvec{Y}$ to obtain feature relevance scores $R_t$ or $R_{k,m}$ for each timepoint or each time-frequency component of the input sample, respectively. The next consideration is how to adequately communicate these scores to users as means of an explanation. In this regard, we propose two explanation formats: the conventional visual approach and an alternative audible approach.

\paragraph{Visual explanations}
In order to provide a visual explanation, we follow the common practice of overlaying the input with a heatmap composed of the relevance values \cite{Jeyakumar2020HowCI}. For the spectrogram this format is very similar to explanations for natural images. For the raw waveform, we employ color-coded timepoint markers based on their respective relevance scores. 
The heatmap is designed as a color map centered at zero, as a relevance score of $R = 0$ indicates a neutral contribution or no impact on the prediction. Positive relevance scores are depicted using red colors, while negative scores are represented by shades of blue.

\paragraph{Audible explanations}
In the domain of audio data, the interpretability of visual explanations may be called into question, as the most natural way for humans to perceive and understand audio is through listening. In the study by Schuller et al. \cite{schuller2021soni}, a roadmap towards XAI for audio is presented, highlighting the importance of providing audible explanations.

Existing XAI methods that provide audible explanation include AudioLIME \cite{haunschmid2020audiolime, melchiorre2021lemons, wullenwber2022coughlime}. This approach  initially performs audio segmentation and source separation to obtain interpretable components. The relevance of these components is then quantified using LIME \cite{ribeiro2016lime} and for the explanation, the top most relevant source segments are played. AudioLIME thus shifts the problem of audible explanations to audio segmentation and source separation. %
In consequence, the final explanation heavily relies on the separation of the signal into interpretable parts. In certain applications, obtaining such a segmentation may not be readily available or straightforward.
We offer an approach that is independent of audio segmentation and source separation algorithms (and thus applicable even in cases where no solutions exists for the specific kind of audio data at hand). This cancels undesired variability of the explanations induced by the specific choice of the source separation algorithm. Instead, we port the basic idea of overlaying input with heatmap to the audible domain, by simply taking the element-wise product between the raw waveform and the heatmap,
\begin{equation} \label{eq:audible_explanation}
     \text{ReLU}(\fvec{R}) \odot \fvec{x} \,.
\end{equation}
In each audible explanation, we can only present either positive or negative relevance. Thus, in \Eqref{eq:audible_explanation}, we mask only the positive relevance. Alternatively, we could do the same for the negative relevance to elucidate what contradicts the prediction.
Currently, our computation of audible explanations is limited to the time domain. This is due to the exclusion of phase information in the spectrograms, making it challenging to directly reconstruct the waveform representation, although it is theoretically possible.

These audible explanations straightforwardly generalize to relevance scores from novel concept-based XAI methods \cite{vielhaben2023multidimensional, achtibat2023attribution},
which is explored in \cite{parekh2022nmf}.

\section{Results}
\label{sec:results}
In this section, we introduce a dataset that can serve as a testbed for the audio AI community. Subsequently, we train two models on distinct input representations, namely the raw waveform and spectrogram. We then proceed to present visual explanations for these models, allowing us to extract high-level classification strategies from the perspective of model developers. Finally, we explore audible explanations and conduct a comparative analysis to assess their interpretability in a human user study, contrasting them with visual explanations and their potential for XAI for end-users.

\subsection{AudioMNIST dataset}
In the computer vision community, the simple MNIST dataset \cite{lecun1998mnist} is still often employed as an initial testbed for model development. Here, we propose an analogous dataset for the audio community and call it \emph{AudioMNIST}.
The AudioMNIST dataset\footnote{Published at: \url{https://github.com/soerenab/AudioMNIST}} consists of 30,000 audio recordings (amounting to a grand total of approx. 9.5 hours of recorded speech) of spoken digits (0-9) in English with 50 repetitions per digit for each of the 60 different speakers. Recordings were collected in quiet offices with a R{\O}DE NT-USB microphone as mono channel signal at a sampling frequency of 48kHz and were saved in 16 bit integer format. In addition to audio recordings, meta information including age (range: 22 - 61 years), sex (12 female / 48 male), origin and accent of all speakers were collected as well. All speakers were informed about the intent of the data collection and have given written declarations of consent for their participation prior to their recording session. The AudioMNIST dataset can be used to benchmark models for different classification tasks of which classification of the spoken digit and the speaker's sex are explored in this paper.

\subsection{Deep spoken digit classifiers}
\label{subsec:audioclassification}
In this section, we implement both a CNN trained on (two-dimensional) spectrogram representations of the audio recordings as well as a CNN for (one-dimensional) raw waveform representations.

\paragraph{Classification based on spectrograms}
\label{subsubsec:classificationbasedonspectrograms}
First, we train a CNN model on the spectrogram representation of the recordings. Its architecture is based on AlexNet \cite{NIPS2012_4824} without normalization layers.

To obtain the spectrograms from the audio recordings, first we downsample them to 8kHz and zero-padded to get an 8000-dimensional vector per recording.
During zero-padding we augment the data by placing the signal at random positions within the vector.
Then, we apply a short-time Fourier transform in \Eqref{eq:stdft} using a Hann window of width 455 and with 420 time points overlap to the signal. This results in spectrogram representations of size $228\times230$ (frequency$\times$time). Next, the spectrograms are cropped to a size of $227\times227$ by discarding the highest frequency bin and the last three time segments. Finally, we convert the amplitude of the cropped spectrograms to decibels and use them as input to the model.

\paragraph{Classification based on raw waveform representations}
\label{subsubsec:classificationbasedonrawwaveforms}
For classification based on raw waveforms, we use the downsampled and zero-padded signals described in \Subsubsecref{subsubsec:classificationbasedonspectrograms} as input to the neural network directly. Here, we design a custom CNN which we refer to as \emph{AudioNet}.
For details of the training protocol of both models and the architecture of AudioNet, we refer to \ref{sec:modeldetails}.

\begin{table}[t!]
  \centering
  \small
  \caption{Mean accuracy $\pm$ standard deviation over data splits for AlexNet and AudioNet on the digit and sex classification tasks of AudioMNIST.}
  \begin{tabular}{cccc}
  Model & Input & \multicolumn{2}{c}{Task} \\

  \cline{3-4}
                         &                        & Digit        & Sex       \\ \hline
  AlexNet                & spectrogram            & $95.82\% \pm 1.49\%$ & $95.87\% \pm 2.85\%$ \\ 
  AudioNet               & waveform               & $92.53\% \pm 2.04\%$ & $91.74\% \pm 8.60\%$ \\ 
  \end{tabular}
  \label{tab:table1}
\end{table}

Model performances are summarized in \Tabref{tab:table1} in terms of means and standard deviations across test folds. Comparisons of model performances may be difficult due to the differences in training parameters and are also not the primary goal of this paper, yet, we note that AlexNet performs consistently superior to AudioNet for both tasks. However, both networks show test set performances well above chance level, i.e., for both tasks the networks discovered discriminant features within the data. The considerably high standard deviation for sex classification of AudioNet originates from a rather consistent misclassification of recordings of a single speaker in one of the test folds.

\subsection{Visual explanations reveal classifier strategies} \label{sec:visual_explanations}
In this section, we visualize LRP relevances for AlexNet and AudioNet. We then derive high-level model classification strategies from the explanations, that we evaluate in sample manipulation experiments.

\begin{figure}[!t]
\centering
  \subfloat[female speaker \\ \phantom{(a)}\centering \textit{digit zero}\label{fig:spectro_0_female_digit}]{
	\includegraphics[width=0.49\linewidth]{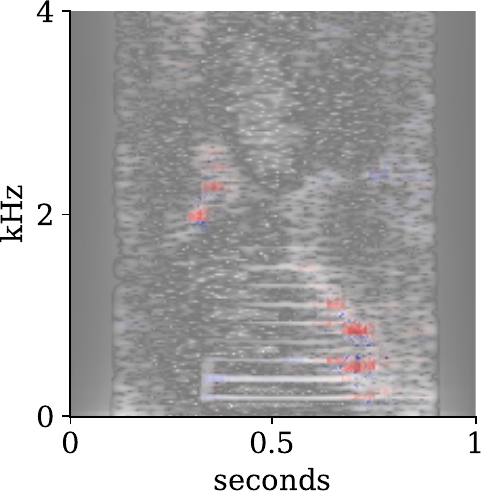}
 } 
  \subfloat[female speaker\\ \phantom{(a)}\centering \textit{digit one}\label{fig:spectro_1_female_digit}]{
	\includegraphics[width=0.49\linewidth]{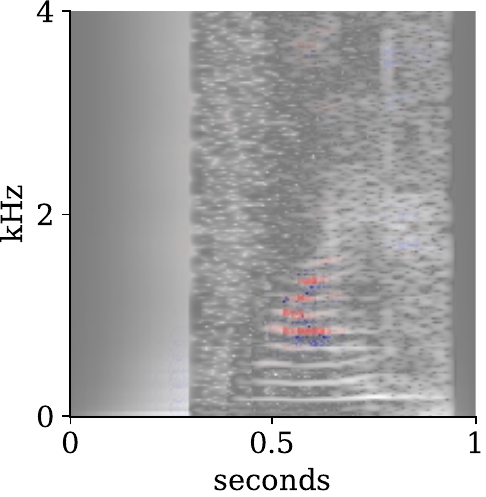}
 }\\
 \subfloat[ \textit{female speaker}\\ \phantom{(a)}\centering digit zero\label{fig:spectro_0_female_gender}]{
	\includegraphics[width=0.49\linewidth]{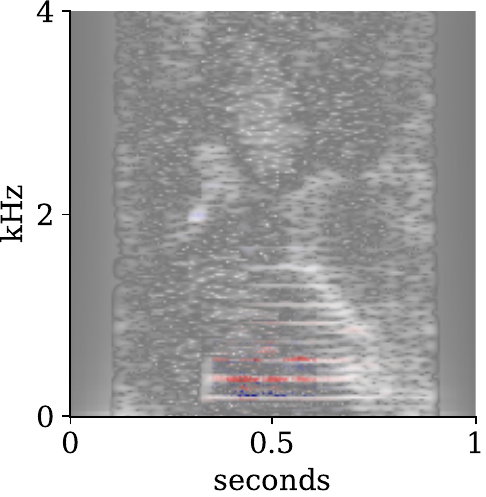}
 }
  \subfloat[ \textit{male speaker}\\ \phantom{(a)}\centering digit zero\label{fig:spectro_0_male_gender}]{
	\includegraphics[width=0.49\linewidth]{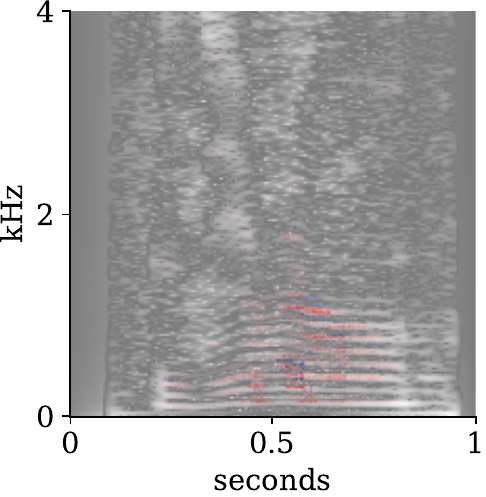}
 }
  \caption{Visual explanations --- spectrograms as input to AlexNet with LRP relevance heatmaps overlayed. 
  \textit{Left, \protect\subref{fig:spectro_0_female_digit} and \protect\subref{fig:spectro_1_female_digit}:} Digit classification.
  \textit{Right, \protect\subref{fig:spectro_0_female_gender} and \protect\subref{fig:spectro_0_male_gender}:} Sex classification.
  Data in \protect\subref{fig:spectro_0_female_digit} and \protect\subref{fig:spectro_0_female_gender} is identical.
  Text in \textit{italics} below the panels indicate the prediction task and explained outcome.
  In all cases, predictions are correct and the true class is explained by LRP.}
\label{fig:spectro}
\end{figure}

\paragraph{Relevance maps for AlexNet}
We compute LRP relevance scores for the AlexNet digit and sex classifier and show exemplary visual explanations based on these scores in \Figref{fig:spectro}.

 The spectrogram in  \Figref{fig:spectro}\protect\subref{fig:spectro_0_female_gender} and \Figref{fig:spectro}\protect\subref{fig:spectro_0_male_gender} correspond to a spoken \textit{zero} by a female and a male speaker, respectively. AlexNet correctly classifies both speakers' biological sex. Most of the relevance distributed in the lower frequency range for both classes. Based on the relevant frequency bands it may be hypothesized that sex classification is based on the fundamental frequency and its immediate harmonics which are in fact known discriminant features for sex in speech \cite{traunmuller1995frequency}.

Spectrograms in \Figref{fig:spectro}\protect\subref{fig:spectro_0_female_digit} and \Figref{fig:spectro}\protect\subref{fig:spectro_1_female_digit} correspond to spoken digits \textit{zero} and \textit{one} from the same female speaker.  AlexNet correctly classifies both spoken digits and the LRP scores reveal that different areas
of the input data appear to be relevant for its decision. However, it is not possible to derive any deeper insights about the classification strategy of the model based on these visual explanations.

The spectrogram in \Figref{fig:spectro}\protect\subref{fig:spectro_0_female_gender} is identical to that in \Figref{fig:spectro}\protect\subref{fig:spectro_0_female_digit}, but the former is overlayed with relevance heat-maps from the sex classifier while the latter shows the heatmap from the digit classifier.
As a sanity check, we confirm that although the input spectrograms in \Figref{fig:spectro}\protect\subref{fig:spectro_0_female_digit} and \Figref{fig:spectro}\protect\subref{fig:spectro_0_female_gender} are identical, the corresponding relevance distributions differ, highlighting the task-dependent feature selection between the digit in and the sex classifier.

\begin{figure*}[!t]
\centering
  \subfloat[][]{
	\includegraphics[width=\textwidth]{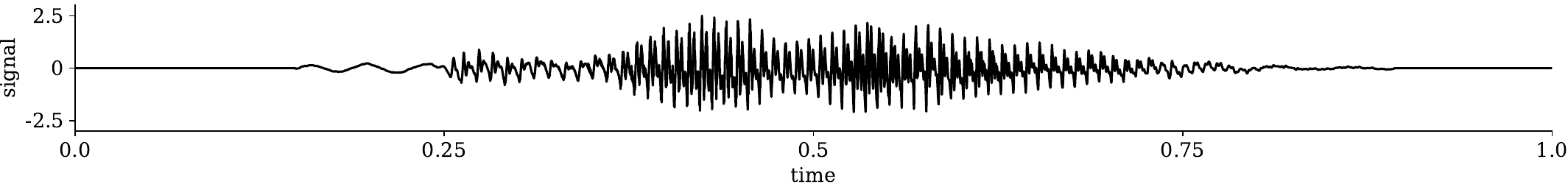}\label{fig:wave_signal}
  }\\ \vspace{-13pt}
  \subfloat[][]{
	\includegraphics[width=\textwidth]{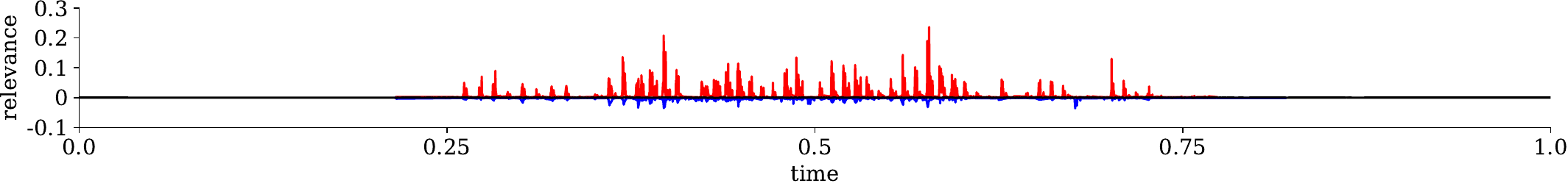}\label{fig:wave_hm}
  }\\ \vspace{-13pt}
  \subfloat[][]{
  	\includegraphics[width=\textwidth]{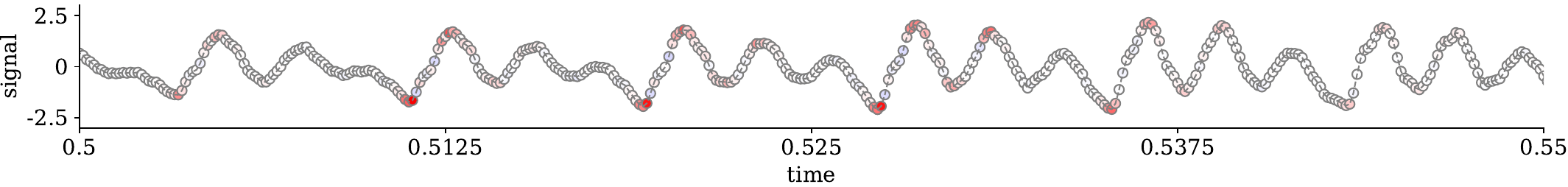}\label{fig:wave_hm_colored}
  }
  \caption{AudioNet correctly classifies the speaker's sex for the waveform in \protect\subref{fig:wave_signal} with associated relevance scores in \protect\subref{fig:wave_hm}.
  Positive relevance in favor of class \textit{male} is colored in red and negative relevance, i.e., relevance in favor of class \textit{female}, is colored in blue.
  A selected range of the waveform from \protect\subref{fig:wave_signal} is again visualized in \protect\subref{fig:wave_hm_colored} with single samples color-coded according to their relevance.
  It appears that mainly samples of large magnitude are relevant for the network's inference.}
\label{fig:wave_heatmap}
\end{figure*}

\paragraph{Relevance maps for AudioNet}
Next, we compute LRP relevance scores for the AudioNet digit and sex classifier that operates on the raw waveforms. 
In \Figref{fig:wave_heatmap}, we show a visual explanation for an exemplary spoken zero from a male speaker for the sex classifier's correct prediction \emph{male}.
Here, we first show the signal and the relevance scores separately in \Figref{fig:wave_heatmap}\protect\subref{fig:wave_signal} and \Figref{fig:wave_heatmap}\protect\subref{fig:wave_hm}. For the time interval from $0.5-0.55$ seconds a zoomed-in segment is provided in \Figref{fig:wave_heatmap}\protect\subref{fig:wave_hm_colored} for an actual visual explanation, where timepoints color-coded according to the relevance score. 
Intuitively plausible, zero relevance falls onto the zero-embedding at the left and right side of the recorded data. Furthermore, from \Figref{fig:wave_heatmap}\protect\subref{fig:wave_hm_colored} it appears that mainly timepoints of large magnitude are relevant for the network's classification decision.

\paragraph{Relevance-guided sample manipulation for AlexNet}
The relevance maps for the AlexNet sex classifier (\Figref{fig:spectro}\protect\subref{fig:spectro_0_female_gender} and \Figref{fig:spectro}\protect\subref{fig:spectro_0_male_gender}) suggest that the sex classifier focuses on differences in the fundamental frequency and subsequent harmonics for feature selection. To investigate this hypothesis the test set was manipulated by up- and down-scaling the frequency-axis of the spectrograms of male and female speakers by a factor of $1.5$ and $0.66$, respectively. Fundamental frequency and spacing between harmonics in the manipulated spectrograms approximately match the original spectrograms of the respectively opposite sex.

After the data has been manipulated as described, the trained network reaches an accuracy of only $20.3\% \pm 12.6\%$ across test splits on the manipulated data, which is well-below chance level for this task.
In other words, identifying sex features via LRP allowed us to successfully perform transformations on the inputs that target the identified features specifically such that the classifier is approximately $80\%$ accurate in predicting the \emph{opposite} sex.

\paragraph{Relevance-guided sample manipulation for AudioNet}
For AudioNet we assess the reliance of the models on features marked as relevant by LRP by an analysis similar to the \textit{pixel-flipping} (or input perturbation) approach introduced in \cite{bach2015pixel, samek2017evaluating}. Specifically, we employ three different strategies to systematically manipulate a fraction of the input signal by setting selected samples to zero. Firstly, as a baseline, (non-zero) samples of the input signal are selected and flipped at random. Secondly, samples of the input are selected with respect to maximal absolute amplitude, e.g the 10\% samples with the highest absolute amplitude are selected, reflecting our (naive) observation from \Figref{fig:wave_heatmap}\protect\subref{fig:wave_hm_colored}. Thirdly, samples are selected according to maximal relevance as attributed by LRP. If the model truly relies on features marked as relevant by LRP, performance should deteriorate for smaller fractions of manipulations in case of LRP-based selection than in case of the other selection strategies.

\begin{figure*}[t!]
\centering
    \subfloat{
      \includegraphics[width=.5\textwidth]{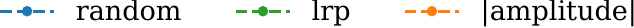}
    }\par\medskip
    \begin{minipage}{0.49\textwidth}
      \centering
      \addtocounter{subfigure}{-1}
      \subfloat[Digit Classification\label{fig:perturbation_digit_setzero_digit}]{
        \includegraphics[width=0.9\linewidth]{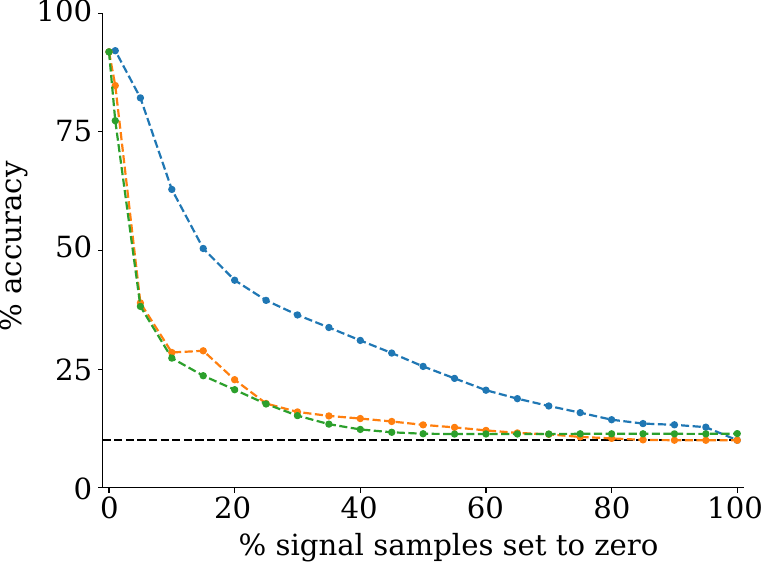}
      }
    \end{minipage}
    \begin{minipage}{0.49\textwidth}
      \centering
      \subfloat[Sex Classification\label{fig:perturbation_digit_setzero_gender}]{
        \includegraphics[width=0.7\linewidth]{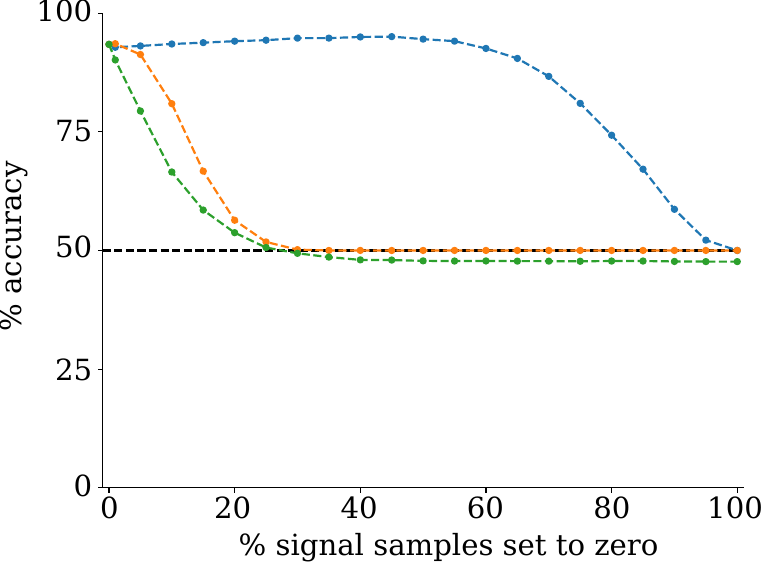}
      }
    \end{minipage}
    \caption{Assessment of classifer reliance on relevant timepoints by flipping timepoints to zero in order of their relevance scores. \protect\subref{fig:perturbation_digit_setzero_digit}: Digit classification. \protect\subref{fig:perturbation_digit_setzero_gender}: Sex classification. Signal samples are either selected randomly (blue), based on their absolute amplitude (orange) or their relevance according to LRP (green). The dashed black line shows the chance level for the respective task.}
    \label{fig:perturbation_digit_setzero}
  \end{figure*}
Model performances for digit and sex classification on manipulated test sets in relation to the fraction of manipulated samples are displayed in \Figref{fig:perturbation_digit_setzero}.
For both tasks, model performances deteriorate for substantially smaller manipulations for LRP-based sample selection as compared to random selection and for slightly smaller manipulations as compared to amplitude-based selections. The effect becomes most apparent for digit classification where a manipulation of 1\% of the signal leads to a deterioration of model accuracy from originally $92.53\%$ to 92\% (random), 85\% (amplitude-based) and 77\% (LRP-based).
In case of sex classification, the network furthermore shows a remarkable robustness towards random manipulations with classification accuracy only starting to decrease when 60\% of the signal has been set to zero, as shown in \Figref{fig:perturbation_digit_setzero}\protect\subref{fig:perturbation_digit_setzero_gender}.
The decline in model performance in both the relevance-based and amplitude-based perturbation procedures supports our hypothesis that the model seems to ground its inference in the high-amplitude parts of the signal. The fact the relevance-based perturbation has a marginally stronger impact on the model however tells us that while our initial hypothesis strikes close to the truth, our interpretation of the model's reasoning based on visual explanations is not exhaustive.

\subsection{Audible explanations surpass visual for\\ interpretability} \label{sec:userstudy}
In the above experiments we investigate the overall model behaviour from a technical point of view, which is a XAI use-case scenario targeted mostly at model developers. However, a purely visual explanation as in \Figref{fig:wave_heatmap} may be insufficient to communicate the model reasoning underlying a single model prediction to the non-expert end user.
Specifically, we investigate the question, which explanation format is the most interpretable to humans: audible or visual explanations. Previous work on XAI for audio shows a combination of both \cite{melchiorre2021lemons, haunschmid2020audiolime}. Here, we show either an audible or visual explanation and compare their interpretability in a human user study, where we measure the human-XAI performance.

\paragraph{Study design}
To compare the interpretability of audible and visual explanations, we ask the user to predict the model prediction based on the explanation, following the method to evaluate the human-XAI performance as suggested in \cite{Hoffman2018MetricsFE}. The outcome is particularly interesting for samples where the model prediction does not match the ground truth label, especially given the high prediction accuracy of the classifier. Our study design is similar yet distinct to the human user study design in \cite{parekh2022nmf}, which compares different audible explanations by asking the user for a subjective judgments how well explanation and model prediction relate. To the best of our knowledge, we are the first to conduct a user study that compares the interpretability of audible and visual explanation formats.
\begin{figure*}[ht]
    \centering
    \includegraphics[width=0.99\linewidth]{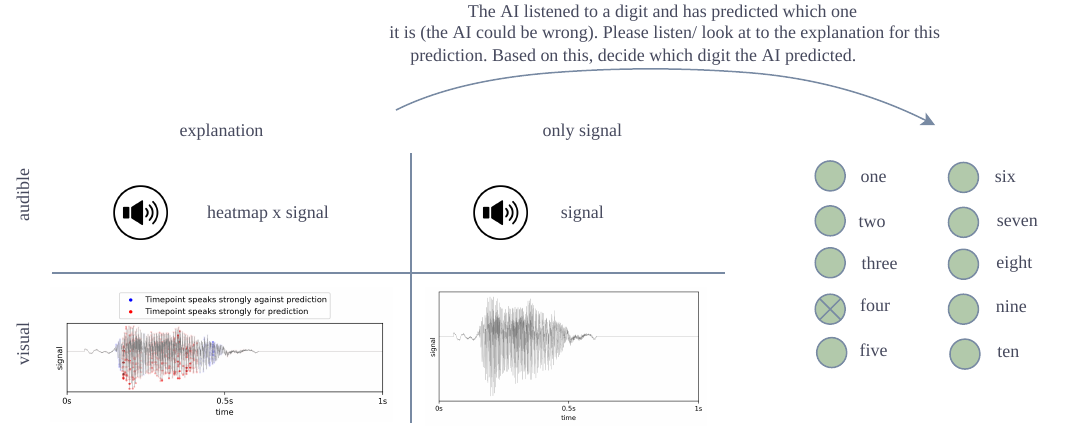}
    \caption{Design of the user study: The user was presented with either a visual or audible explanation. As a baseline we present faux explanations that entail only the signal itself. The user was asked to predict the model prediction based on the explanation.}
    \label{fig:userstudy_design}
\end{figure*}
We compute LRP relevance scores for the AudioNet digit classifier trained on the raw waveforms.\footnote{The user study is based in a reconstruction if the original model in \Secref{subsec:audioclassification}, which originate from an earlier preprint of this work \cite{becker2018earlyaudiomnist}. The model weights and the performance of the reconstruction used in this section slightly deviate from those of the original model.} We choose the digit task, because of its higher complexity compared to sex classification. Further, we focus on explanations for the waveform model because in this representation, relevance can directly be made audible.
In summary, we conduct a comparison between visual explanations which consists of heatmaps overlaying the waveform (as shown in \Figref{fig:spectro} and \Figref{fig:wave_heatmap}), and audible explanations based on \Eqref{eq:audible_explanation}.
As a baseline, we additionally present the user with faux 'explanations' that entail solely the signal itself. Overall, we present both the modulated or overlayed signal with relevance scores as well as solely the signal, for both the audible and visual explanation formats. The study design is visualized in \Figref{fig:userstudy_design}. We choose 10 random samples where the model prediction is correct and 10 random samples where the model is predicting incorrectly.

\paragraph{Data acquisition}
We asked 40 subjects from the authors' research department to predict the model's predictions for the 20 randomly chosen samples and all 4 explanation modes described above, thus each subject answers 80 questions. We also collect meta-information about gender (7.5\% diverse, 15\% female, 77.5\%male), previous experience with XAI (50\% high - researcher in the field,  42.5\% medium - had exposure to XAI, 7.5\% low - roughly knows what XAI is, 0\% zero - never heard of XAI before) and subjectively assessed hearing capabilities (32.5\% very good, 42.5\% good, 25\% medium, 0\% low, 0\% very low). Test subjects gave their informed written consent to participate in the study and to the data acquisition and processing.  A fast-track self-assessment of the study had resulted in a positive evaluation from the ethics commission of the Fraunhofer Heinrich Hertz Institute.

\paragraph{Evaluation}
\begin{figure}[ht!]
    \centering
    \begin{minipage}{0.35\textwidth}
    \subfloat[Incorrect model classification\label{fig:userstudyresults_incorrect}]{
    \includegraphics[width=\linewidth]{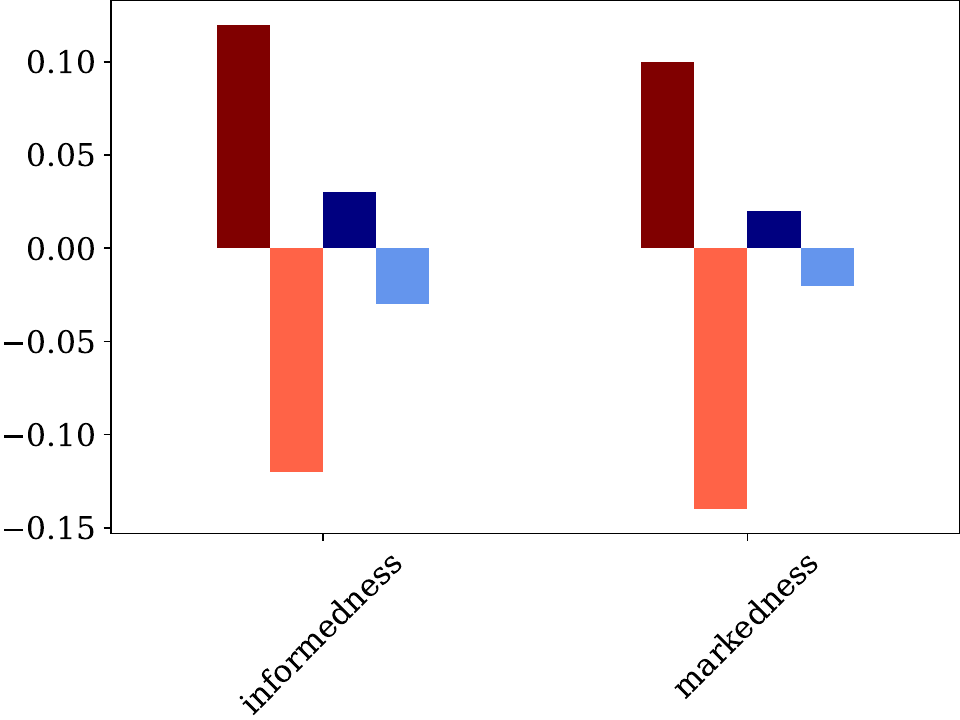}}
    \end{minipage}
    \begin{minipage}{0.35\textwidth}
    \subfloat[Correct model classification\label{fig:userstudyresults_correct}]{
    \includegraphics[width=\linewidth]{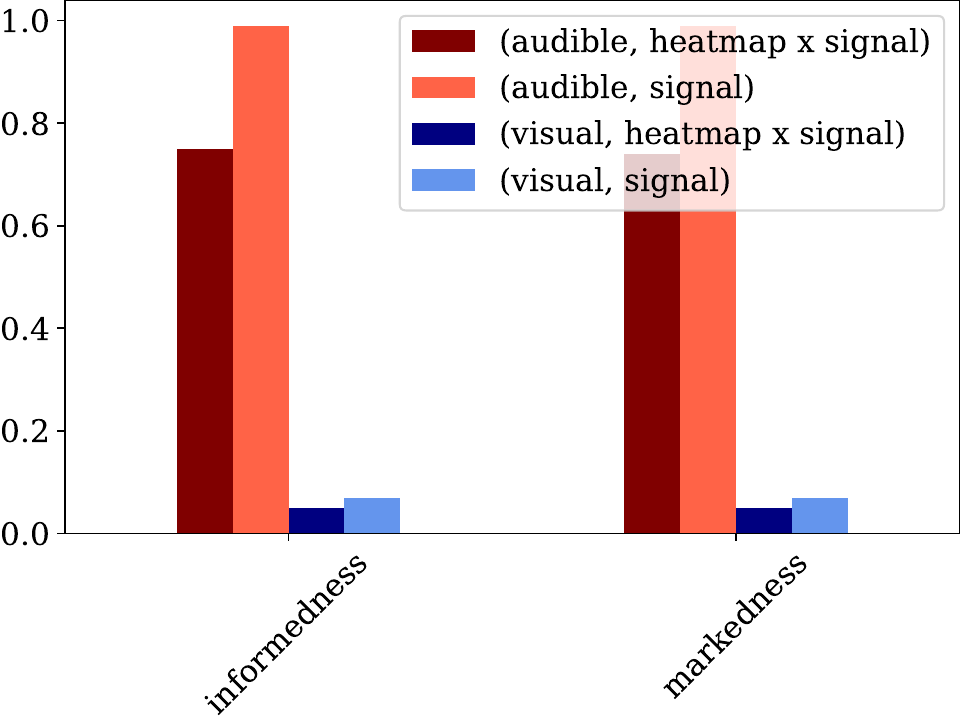}}
    \end{minipage}
    \caption{\protect\subref{fig:userstudyresults_incorrect}: User performance on incorrectly predicted samples based on the different explanation formats. \protect\subref{fig:userstudyresults_correct}: User performance on correctly predicted samples.}
    \label{fig:userstudy_results}
\end{figure}

In in \Figref{fig:userstudy_results}, we show the user's performance in predicting the model's predictions based on the explanations, where we report informedness and markedness for the multi-class case \cite{powers2011evaluation}. Informedness for each class is defined as $TP/P - FP/N$ and measures how \emph{informed} the user is about the positive and negative model predictions for this class based on the explanation. Markedness for each class in $TP/(TP+FP) - FN/(TN+FN)$ and measures the \emph{trustworthiness} of the user's prediction of positive and negative model preditions for this class. Here, \emph{TP, FP, TN, FN} denote true and false positive and negative predictions, respectively. Values for both metrics range from -1 to +1, where positive values imply that the user is informed correctly by the explanation and their prediction can be trusted and negative values imply that the user is informed incorrectly and that it can be trusted that their prediction is wrong. 

First, we evaluate the case where the model prediction does not match the true digit, see \Figref{fig:userstudyresults_incorrect}. We find that audible explanations show a markedly greater informedness and markedness than their visual counterpart. However, the values of 0.12 and 0.1 indicate that there is still room for improvement in terms of the interpretability of audible explanations. This can be achieved by employing innovative concept-based methods that have demonstrated improved interpretability in computer vision applications \cite{achtibat2023attribution}. As expected, the baseline containing only the signal show negative informendess and markedness, as the user is informed incorrectly about the model prediction and it can be trusted that their prediction is wrong.

Second, for the samples where the model classified the digits correctly, as expected the user's prediction performance is higher for all explanation formats than for the incorreclty classified samples, see \Figref{fig:userstudyresults_correct}. Further, both informedness and markedness are higher for the audible signal than for the actual audible explanation. This is natural as it is possible that the model's classification strategy deviates from the user's classifiaction strategy. To illustrate this, across all digits classes, both informedness and markedness have the lowest value for the samples correctly classified as a 'nine' by the model with a value of 0.33. Here, 33\% users predicted that the model classified the digit as a 'nine' and 32\% predicted that the model classified it as a 'five'. In the explanation, only the common syllable, the 'i' is audible. Like for the samples incorrectly classified by the model, both informedness and markedness are much lower for the visual explanations than for the audible ones. Interestingly, the comparison between explanation and signal only follows the same trend as for their audible counterparts.

We conclude that audible explanations for audio classifiers exhibit a higher level of interpretability compared to visual explanations for human users. This highlights the importance of the presentation of the explanation, beyond the mere computation of raw relevance values. Furthermore, it emphasizes that the optimal format of explanations may vary across different applications. In the context of audio applications, the superior interpretability of audible explanations is expected, considering that listening is the innate way for humans to perceive audio signals.
As proposed above, to further improve the interpretability of audible explanations, concept-based approaches \cite{vielhaben2023multidimensional,achtibat2023attribution}, that put the model prediction for a single sample into context with the model reasoning over the entire dataset, could be leveraged and made audible, similar to the work in \cite{parekh2022nmf}.

\section{Conclusion}
\label{sec:conclusion}
The need for interpretable model decisions is increasingly evident in various machine learning applications. While existing research has primarily focused on explaining image classifiers, there is a dearth of studies in interpreting audio classification models.  To foster open research in this direction we provide a novel Open Source dataset of spoken digits in English as raw waveform recordings. Further, we demonstrated that LRP is a suitable XAI method for explaining neural networks for audio classification.
By employing visual explanations based on LRP relevances, we have successfully derived high-level classification strategies from the perspective of model developers. Most notably, we have introduced audible explanations that align with the established framework for explanation presentation in computer vision. Through a user study, we have conclusively shown that audible explanations exhibit superior interpretability compared to visual explanations for the classification of individual audio signals by the model.

In future work we aim to apply LRP to more complex audio datasets to gain a deeper insight into classification decisions of deep neural networks in this domain. 
Further, we aim to improve the interpreatbility of audible explanations, by using concept-based XAI methods as studied in \cite{vielhaben2023multidimensional, achtibat2023attribution}.

\section*{Acknowledgement}
WS and KRM were supported by the German Ministry for Education and Research (BMBF)
under grants 01IS14013A-E, 01GQ1115, 01GQ0850, 01IS18056A, \\ 01IS18025A and 01IS18037A.
WS, SL and JV received funding from the European Union’s Horizon 2020 research and innovation programme under grant iToBoS (grant no.\ 965221), from the European Union’s Horizon Europe research and innovation programme (EU Horizon Europe) as grant TEMA (grant no.\ 101093003), and the state of Berlin within the
innovation support program ProFIT (IBB) as grant BerDiBa (grant no.\ 10174498).
WS was further supported by the German Research Foundation (ref.\ DFG KI-FOR 5363).
KRM was also supported by the Information \& Communications Technology Planning \& Evaluation (IITP) grant funded by the Korea government (grant no.\ 2017-0-001779), as well as by the Research Training Group “Differential Equation- and Data-driven Models in Life Sciences and Fluid Dynamics
(DAEDALUS)” (GRK 2433) and Grant Math+, EXC 2046/1, Project ID 390685689 both funded
by the German Research Foundation (DFG).

\bibliographystyle{elsarticle-num}
\bibliography{bibliography}


\appendix

\section{Model details} \label{sec:modeldetails}
We provide some further details on the architecture and training protocols for the audio classification models in \Secref{subsec:audioclassification}.

\paragraph{AudioNet architecture}
AudioNet consists of 9 weight layers that are organized in series as follows\footnote{Layer naming pattern examples: conv3-100 -- conv layer with 3x1 sized kernels and 100 output channels. FC-1024 -- fully connected layer with 1024 output neurons}: conv3-100, maxpool2, conv3-64, maxpool2, conv3-128, maxpool2, conv3-128, maxpool2, conv3-128, maxpool2, conv3-128, maxpool2, FC-1024, FC-512, FC-10 (digit classification) or FC-2 (sex classification). All convolutional layers employ a stride of 1 and are activated via ReLU nonlinearities. Max-pooling layers employ stride 2.

\paragraph{Dataset splits}
For digit classification, the dataset was divided by speaker into five disjoint splits each containing data of 12 speakers, i.e., 6,000 spectrograms per split. In a five-fold cross-validation, three of the splits were merged to a training set while the other two splits respectively served as validation and test set. In a final, fold-dependent preprocessing step the element-wise mean of the training set was subtracted from all spectrograms.

For sex classification, the dataset was reduced to the 12 female speakers and 12 randomly selected male speakers. These 24 speakers were divided by speaker into four disjoint splits each containing data from three female and three male speakers, i.e., 3,000 spectrograms per split. In a four-fold cross-validation, two of the splits were merged to a training set while the other two splits served as validation and test set. All other preprocessing steps and network training parameters were identical to the task of digit classification.

\paragraph{Model training}
For both the sex and digit classification task, AlexNet was trained with Stochastic Gradient Descent for 10,000 optimization steps at a batchsize of 100 spectrograms. The initial learning rate of 0.001 was reduced by a factor of 0.5 every 2,500 optimization steps, momentum was kept constant at 0.9 throughout training and gradients were clipped at a magnitude of 5.

 In case of digit classification, AudioNet was trained with Stochastic Gradient Descent with a batch size of 100 and constant momentum of 0.9 for 50,000 optimization steps with an initial learning rate of 0.0001 which was lowered every 10,000 steps by a factor of 0.5. In case of sex classification, training consisted of only 10,000 optimization steps where the learning rate is reduced after 5,000 steps.

\end{document}